\documentclass[structabstract]{aa}
\usepackage{graphicx}
\usepackage{txfonts}
\usepackage{color}
\begin{document}

\title{A spectroscopic survey of faint, high-galactic latitude red clump stars. II. The medium resolution sample}
\titlerunning{A spectroscopic survey of faint, high-galactic latitude red clump stars. II}

\author{T. Saguner \inst{1,2}, U. Munari\inst{1}, M. Fiorucci \inst{1} and A. Vallenari\inst{1}}
\authorrunning{T. Saguner et al.}

\institute {$^1$INAF-OAPD-Osservatorio Astronomico di Padova, via dell'Osservatorio 8,I-36012 Asiago (VI), Italy \\ 
	    $^2$University Of Padova, Department Of Astronomy, vicolo dell'Osservatorio 3, I-35122 Padova (PD), Italy \\    
	   }
\date{Received      , 2010; accepted      , 2010}

\abstract
{}
{The goal of our survey is to provide accurate and multi-epoch radial
velocities, atmospheric parameters ($T_{\rm eff}$, $\log g$ and [M/H]),
distances, and space velocities of faint red clump stars.}
{We recorded high signal-to-noise (S/N$\geq$200) spectra of RC stars
over the 4750-5950~\AA\ range at a resolving power 5500. The target stars
are distributed across the great circle of the celestial equator. Radial
velocities were obtained via cross-correlation with IAU radial velocity
standards. Atmospheric parameters were derived via $\chi^2$ fit to a synthetic
spectral library. A large number of RC stars from other surveys were
re-observed to check the consistency of our results and the absence of offsets
and trends.}
{A total of 245 RC stars were observed (60 of them with a second
epoch observation separated in time by about three months), and the results
are presented in an output catalog.  None of them is already present in
other surveys of RC stars.  In addition to astrometric and
photometric support data from external sources, the catalog provides radial
velocities (accuracy $\sigma$(RV)=1.3 km s$^{-1}$), atmospheric parameters
($\sigma(T_{\rm eff})$=88 K, $\sigma(\log g)$=0.38 dex and
$\sigma$([M/H])=0.17 dex), spectro-photometric distances, (X,Y,Z)
galacto-centric positions and (U,V,W) space velocities.}
{} 

\maketitle

\section{Introduction}

The Red Clump (RC) is composed by low mass stars in the stage of central
helium burning, following He ignition in an electron-degenerate core
(Girardi 1999).  They display properties that make them a primary tool to
investigate Galactic structure and kinematics: \textit{(i)} their absolute
magnitude shows minimal dispersion at optical and infrared wavelengths,
\textit{(ii)} they are intrinsically bright, and thus observable throughout
most of the Galaxy, \textit{(iii)} in magnitude-limited surveys they count
for a fairly large fraction of observed targets, \textit{(iv)} their
spectral types, ranging mainly from G8III to K2III, make them ideal stars to
measure accurate radial velocities and atmospheric chemical abundances. 
Examples of recent applications of RC stars to Galaxy investigations are,
among countless more, the peculiarities of Galactic rotation (Rybka et al. 
2008), the stellar bar in the inner Galaxy (Cabrera-Lavers et al.  2007),
the Galactic Bulge (Nataf et al.  2010), the vertical distribution of disk
stars in terms of kinematic and metallicity (Soubiran et al.  2003), the
surface mass density in the Galactic plane (Siebert et al.  2003), the
origin of the Thick disk (Ruchti et al.  2010), the surface mass density in
the Galactic disk (Bienayme et al.  2006) and age-metallicity relation
(AMR), age-velocity relation (AVR) (Soubiran et al.  2008), tidal streams in
solar neighborhood (Famaey et al.  2005, Antoja et al.  2008), Galactic
substructures (Correnti et al.  2010, Law et al.  2010).  The large
proportion of RC stars observed by the ongoing RAVE survey (Steinmetz et al. 
2006, Zwitter et al.  2008) and the accurate distances derived for them
(Zwitter et al.  2010) support a great potential of the RAVE data base in
progressing towards a better understanding of how the Galaxy formed,
structured and evolved (Freeman \& Bland-Hawthorn 2002, Siebert et al.  
2008, Veltz et al.  2008, Kiss et al.  2010).

This paper is an extension to fainter magnitudes of the spectroscopic survey
of RC stars of Valentini \& Munari (2010), hereafter named Paper I. 
Accurate radial velocities, atmospheric parameters ($T_{\rm eff}$, $\log g$,
[M/H]), distances and space velocities have been obtained for 245 RC stars
distributed along the great circle of celestial equator, 60 of them
re-observed at a second epoch.  These data are presented in a catalog
together with photometric and astrometric support information from external
sources. 

The application to Galaxy investigations of the results obtained in this
study and in Paper I will be the topic of a forthcoming paper.

\section{Target selection criteria}

To ensure the highest homogeneity within the present program, we adopted
exactly the same target selection criteria as in Paper I, the only
difference being an average magnitude of the target stars $\sim$1 mag
fainter than in Paper I and consequently a $\sim$50\% larger distance.

To summarize, the selection criteria are: \textit{(i)} a star of spectral
type between G8III and K2III as classified by the Michigan Project (Houk and
Swift 1999), with \textit{(ii)} a high accuracy in the spectral type
(quality index $\leq$2), and \textit{(iii)} a blank spectroscopic duplicity
index.  The stars must be \textit{(iv)} uniformly distributed in right
ascension within $\pm$6$^\circ$ of the celestial equator (thus along a great
circle on the sky), and at \textit{(v)} Galactic latitude
$|$$b$$|$$\geq$25$^\circ$ (to avoid the increasing reddening close to the plane of the
Galaxy).  In addition, the target stars must \textit{(vi)} be valid entries
in both the Hipparcos\footnote{All selection criteria involving Hipparcos
information refer to the original catalog published by ESA in 1997, and not
necessarily its revision by van Leeuwen (2007).} and Tycho-2 catalogs,
\textit{(vii)} have non-negative Hipparcos parallaxes, \textit{(viii)}
avoid any other Hipparcos or Tycho-2 star closer than 10 arcsec on the sky,
\textit{(ix)} be confined within the magnitude range 7.8$\leq$$V_{\rm
Tycho−2}$$\leq$9.5 (for Paper I it was 6.8$\leq$$V_{\rm Tycho−2}$$\leq$8.1),
\textit{(x)} have an absolute magnitude (from Hipparcos parallax)
incompatible with either luminosity classes V or I, \textit{(xi)} have a
blank photometric variability flag and a blank duplicity index in the
Hipparcos catalog.  Finally, the target stars must \textit{(xii)} be present
in the 2MASS survey, \textit{(xiii)} be absent in the N\"{o}rdstrom et al. 
(2004) Geneva-Copenhagen survey of dwarfs in the solar neighborhood, and
\textit{(xiv)} be absent in the radial velocity survey of giant stars by
Famaey et al.  (2005). Given the small overlap in magnitude with 
Paper I, we added a final selection criterion that \textit{(xv)} no target
from Paper I is re-observed here, to increase the total number of surveyed RC
stars.

These selection criteria returned a total sample of 500 possible target stars.

\begin{figure}[htp]
\center
\includegraphics[totalheight=0.36\textheight,angle=270]{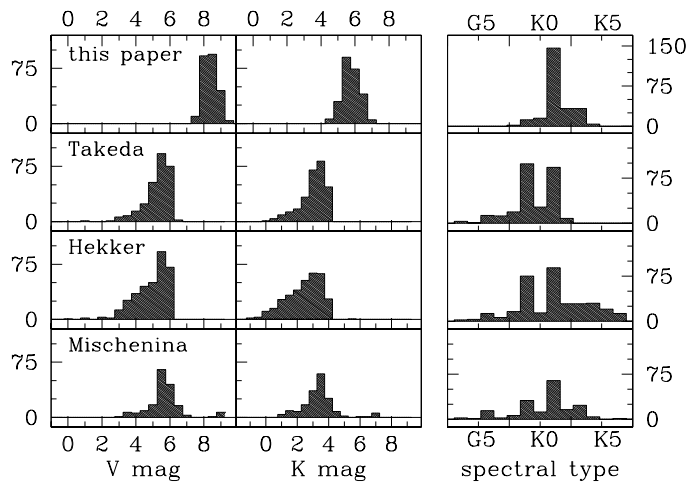}
\caption{Distribution in $V$ and $K$ band brightness, and in spectral type
of our target stars and the RC surveyed by Takeda et al.  (2008), Hekker \&
Melendez (2007), Mishenina et al.  (2006).}
\end{figure}

\section{Program stars}

We have observed 245 target stars, 60 of them at two distinct epochs. 
Figure~1 compares their brightness distribution in $V$, $K$, and spectral
type with those of other surveys of RC stars.  The comparison with the
program stars of Paper I is carried out in Fig.~2.

As for Paper I, we have re-observed a number of RC stars from other surveys
with the same instrument set-up and reduction/analysis procedures.  We
re-observed 87 RC stars so selected: 47 from Hekker \& Melendez
(2007), 34 from Takeda et al. (2008) and 6 from Soubiran
\& Girard (2005).  In addition, 15 IAU radial velocity standards
(most of them RC stars) were observed too.  This allowed us to check for
zero-point offsets in the atmospheric parameters and radial velocity.  These
observations, supplemented by the results of repeated observations of our
target stars, provided a consistent estimate of the errors associated to our
measurements.

\section{Observations and data reduction}

The spectra were obtained from 2008 June to 2010 June with the 1.22m
telescope + B\&C spectrograph operated in Asiago by the Department of
Astronomy of the University of Padova.  The CCD camera was an ANDOR iDus
440A, equipped with an EEV 42-10BU back-illuminated chip, 2048$\times$512
pixels of 13.5~$\mu$m size.  A 1200 ln/mm grating provided a scale of 0.61
\AA/pix and the recorded wavelength range extended from 4750 to 5950 \AA. 
A slit width of 2 arcsec provided a resolving power close to 5500, or a
resolution of about 1.0 \AA.  The exposure time was kept fixed to 600 sec
for all target stars, which provided a S/N$\geq$200 on the final extracted
spectra at the faint end of the magnitude distribution of our program stars. 
For the brighter template RC stars selected from literature and the IAU radial
velocity stars, shorter exposure times were adopted to avoid approaching CCD
saturation levels. All target stars were observed within 1 hour of their 
culmination time.

The 4750 to 5950 \AA\ wavelength range includes the H$\beta$, Mg$b$ triplet
and NaI D$_{1,2}$ doublet regions.  Preliminary to initiating the observations of
the target RC stars, we experimented with bluer and redder wavelength
ranges by extensively observing the same selected template RC stars.  Bluer
ranges suffered from a marked reduction in the S/N of extracted spectra
(lower target brightness and overall instrumental throughput) and a more
difficult continuum normalization owing to increased density of absorption
lines. The lower density of absorption lines of redder wavelenth ranges
offered the advantage of an easier continuum normalization, but at the expense
of a lower diagnostic content for measurement of atmospheric parameters
and radial velocities. If the two effects roughly balanced a marked
disadvantage of redder wavelenth ranges is the presence of telluric absorption
lines and bands which significantly alter the response of $\chi^2$ fits to
synthetic spectra. Eventually, the 4750 to 5950 \AA\ wavelength range was
selected as the best-performing interval.

The data reduction was carried out in \textit{IRAF}, following in detail the
standard procedures described in the manual by Zwitter \& Munari (2000),
which involved correction for bias (from overscan regions), dark and
flat-field frames (from exposure on the dome), sky-background subtraction
(our long-slit mode extends for 1 arcmin on the sky), scattered light
removal, wavelength calibration (from Fe-Ar comparison lamp frames
exposed immediately before and after the science exposure, with the
telescope tracking the star) and heliocentric correction.

At least three IAU radial velocity standards were observed at very high S/N
during each night to allow the derivation of the radial velocity of program
stars via cross-correlation techniques.  

\begin{table}
\caption{Comparison for the program 15 IAU RV standards between the
         tabulated heliocentric radial velocity and that measured in our
         spectra.}
\begin{center}
\begin{tabular}{l l r c c r c}
\hline
\multicolumn{7}{c}{}\\
Star   & Spc.T.    & \multicolumn{2}{c}{Literature}  & & \multicolumn{2}{c}{Cross-Correlation}     \\ \cline{3-4} \cline{6-7} 
       &           & RV$_\odot$  & err. & &   RV$_\odot$ & $\sigma_{\rm RV}$  \\ 
\multicolumn{7}{c}{}\\   
HD 4388   & K3 III & $-$28.3   &  0.6  &  &    $-$27.4   &  2.1   \\         
HD 62509  & K0 III &  03.3     &  0.1  &  &     02.3     &  2.7   \\
HD 132737 & K0 III & $-$24.1   &  0.3  &  &    $-$25.0   &  1.6   \\
HD 136202 & F8 III &  53.5     &  0.2  &  &     48.6     &  1.9   \\
HD 144579 & G8 IV  & $-$60.0   &  0.3  &  &    $-$62.5   &  0.8   \\
HD 145001 & G5 III & $-$09.5   &  0.2  &  &    $-$10.5   &  1.2   \\
HD 154417 & F8 IV  & $-$17.4   &  0.3  &  &    $-$15.3   &  1.5   \\
HD 161096 & K2 III & $-$12.0   &  0.1  &  &    $-$13.6   &  0.9   \\
HD 182572 & G7 IV  & $-$100.5  &  0.5  &  &    $-$102.1  &  1.1   \\
HD 187691 & F8 V   &  00.1     &  0.3  &  &     02.6     &  0.8   \\
HD 194071 & G8 III & $-$09.8   &  0.1  &  &    $-$06.9   &  0.5   \\
HD 212943 & K0 III &  54.3     &  0.3  &  &     55.9     &  1.9   \\
HD 213014 & G9 III & $-$39.7   &  0.0  &  &    $-$37.4   &  1.7   \\
HD 213947 & K2     &  16.7     &  0.3  &  &     17.5     &  1.3   \\
BD+28.3402& F7 V   & $-$36.6   &  0.5  &  &    $-$36.5   &  0.4   \\
\multicolumn{7}{c}{}\\
\hline      
\end{tabular}
\end{center} 
\end{table}

\begin{figure*}[htp]
\center
\includegraphics[totalheight=0.72\textheight,angle=270]{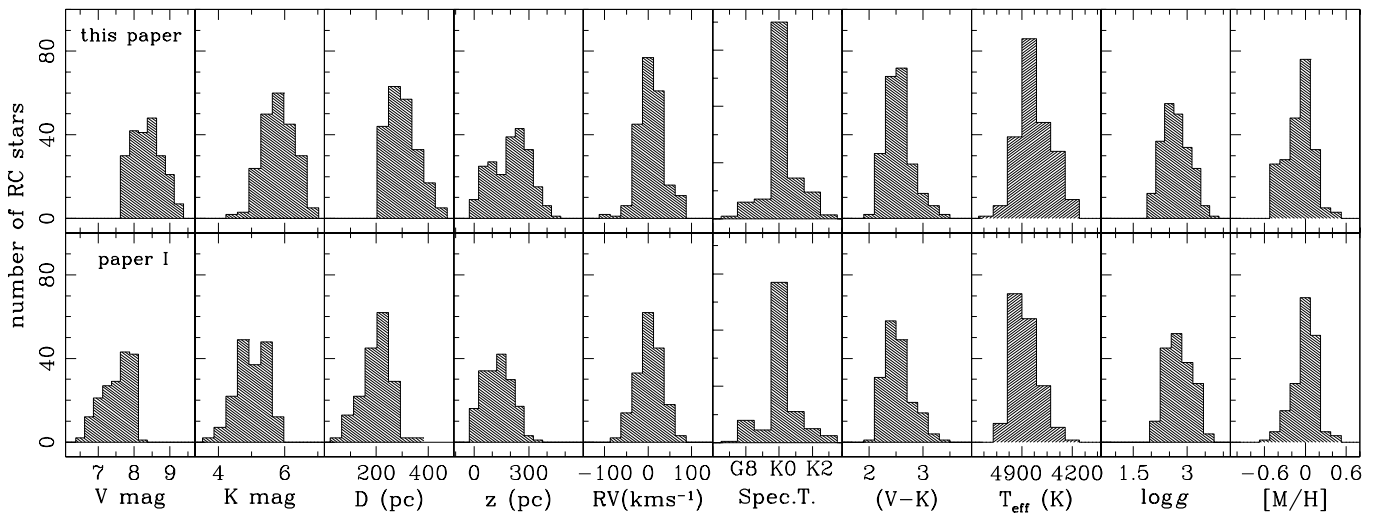}
\caption{Comparison of the distribution in various parameters of the
stars investigated here and in Paper I.}
\end{figure*}

\subsection{Continuum Normalization}

The determination of radial velocities via cross-correlation, and in
particular the derivation of atmospheric parameters via $\chi^2$ fitting,
critically depend on the accuracy of continuum normalization. To ensure
the highest quality, control, and homogeneity, the continuum normalization
of all spectra was carried out manually as a three-step procedure.

First of all, a lot of care was put in training the operator's human
response.  To this aim, we used the synthetic spectral library of Munari et
al.  (2005).  It covers the whole parameter space of the HR diagram and
comes in many different resolving powers, one of which exactly matches that
of our spectra.  The synthetic library provides the same spectra computed
both as absolute fluxes and as continuum normalized.  About 100 synthetic
spectra similar to those of RC stars and of the absolute-flux type where
selected from the synthetic library.  They were multiplied by randomly
selected, arbitrary instrumental response functions, and noise was added to
simulate the S/N$\geq$200 that characterizes our observed spectra.  The
resulting spectra were then manually normalized with the IRAF task {\sc
Continuum}.  To close the loop, the result of the normalization was compared
with the synthetic spectrum originally computed as continuum normalized.  The
comparison easily revealed where improvements in the human decision chain
were necessary, and by a tedious - but also effective - trial-and-error
procedure, a relevant experience and confidence in the results was built. 
On average, the best results were obtained with a Legendre fitting
polynomial of 5$^{th}$ order, a high-rejection threshold of 0.0 and a
low-rejection threshold of 1.35$\times$sigma.

The second step was to apply this human experience to the observed
spectra, normalizing them manually and individually with the same IRAF task,
and whenever appropriate adopting the same fitting polynomials and associated
parameters.  In some cases, a different degree for the polynomial, or masking
of some portion of the spectrum, or changes of other parameters were
necessary for an optimal result, depending on the specific characteristic of
individual observed spectra.

The third and final step was to automatically renormalize with the same IRAF
task {\sc Continuum} and associated parameters the whole Munari et al. 
(2005) library of synthetic spectra, so they would probably speak the same language
as the observed ones.  Before running the $\chi^2$ a final refinement was
applied by imposing that the geometric mean of each synthetic spectrum
matched that of the observed one (this forced the mean level of the
continuum in the observed and synthetic spectra to coincide).

\section{Data quality control}

The wavelength range covered by our spectra includes four appreciably
intense night-sky and city-lights lines, namely [OI] 5577.333, and HgI
5460.734, 5769.579, 5790.643 \AA.  We measured them on all recorded spectra
to check the wavelength calibration from their radial velocity and the
spectrograph focusing from their sharpness.  During data analysis we noticed
that the spectra characterized by a FWHM$\leq$2.0~\AA\ for the night-sky
lines, invariably returned the radial velocity of the night sky lines as
$|$RV$|$$<$1.8 km s$^{-1}$, independently of the FWHM.  The spectra with
FWHM$>$2.0~\AA, showed instead a larger dispersion of the radial velocity of
the night sky lines, which increased with increasing FWHM.

The same dividing threshold affected the accuracy of radial velocities and
atmospheric parameters obtained for the target stars.  Comparing the results
of repeated observations of the same targets, spectra characterized by a
FWHM$\leq$2.0~\AA\ for the night-sky lines showed a well behaving, sharp, and
Gaussian distribution of the differences (see Figures~3 and 5, and text
below), which could not be further improved by a stricter limit imposed
on the FWHM of the night-sky lines.  Conversely, spectra with
FWHM$>$2.0~\AA\ showed larger dispersion in the differences between
1$^{st}$ and 2$^{nd}$ observations, clearly tracing a different and looser
statistical population.

Because we are interested only in the best products our instrumentation could
deliver, we ignored all observations characterized by a FWHM$>$2.0~\AA\
for the night-sky emission lines, and they will not be further considered in
this paper. This meant reducing by 43 objects the number of targets inserted
in the paper's output catalog, which in its final version includes 245 RC
target stars.

\section{Radial velocities}

The radial velocity for our target stars was obtained via cross-correlation
against a set of RC stars that were also IAU radial velocity standard stars.
At least three such standards were observed each night together with the target
stars. The radial velocity of the latter was obtained as the mean
of the cross-correlation results against all standards observed that night.

The cross-correlation was performed in IRAF with the task {\sc fxcor}. About
20~\AA\ were masked at both ends of the spectra, because there the accuracy
of wavelength calibration and continuum normalization degrades for obvious
reasons.

\subsection{Tests on RV standards}

To test the accuracy of our radial velocities, we selected the sample of 15
bright IAU radial velocity standards listed in Table~1.  We observed them
during four different nights.  For each night and for each standard the
radial velocity was derived by cross-correlation against all other standards
observed that same night.  The last two columns of Table~1 given the mean
value and the dispersion of the individual values.

The median dispersion of these measurement is 1.36 km~s$^{-1}$. Considering
that the mean uncertainty of the tabulated radial velocities is 0.27
km~s$^{-1}$, we may conclude that the accuracy of our radial velocities is
1.3 km~s$^{-1}$.

The zero point of our radial velocities is not biased. In fact, the mean
difference between observed and published radial velocity for the 15
standards in Table~1 is only +0.04 km~s$^{-1}$, with an error of the mean of
0.55 and a dispersion of 2.1 km~s$^{-1}$.

\begin{figure}
\center
\includegraphics[totalheight=0.36\textheight,angle=270]{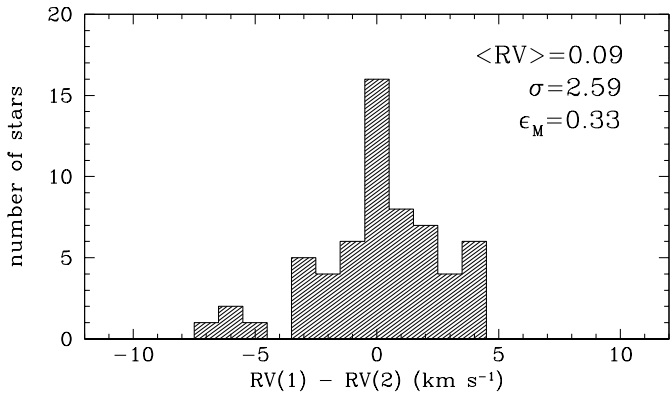}
\caption{Distribution of the differences in radial velocity between repeated
observations.}
\end{figure}

\begin{table}
\caption{Comparison between the atmospheric parameters obtained with
the $\chi^2$ method and those derived by Hekker and Melendez (2007)}
\begin{center}
\begin{tabular}{@{~~~}l@{~~~}c@{~~~~}c@{~~~}c@{~~~}r@{~~~}c@{~~~}c@{~~~}c@{~~~}r@{~~~}}
\hline
\multicolumn{9}{c}{}\\
 HD    & Spc.T.  & \multicolumn{4}{c}{Hekker \& Melendez} &  \multicolumn{3}{c}{$\chi^2$}  \\ \cline{3-5} \cline{7-9} 
       &           & $T_{\rm eff}$ & log\textit{g} & [M/H] &    & $T_{\rm eff}$ & log\textit{g} & [M/H] \\ 
       &           &     (K)       &    (dex)      & (dex) &    &     (K)       &   (dex)       & (dex) \\
\multicolumn{9}{c}{}\\
  3807 &  K0 III &  4625 &  2.30  &  $-$0.44   &  &   4526   &  1.80  & $-$0.46 \\
  4627 &  G8 III &  4599 &  2.05  &  $-$0.25   &  &   4664   &  2.42  & $-$0.04 \\
  6186 &  K0 III &  4829 &  2.30  &  $-$0.24   &  &   4845   &  2.55  & $-$0.50 \\
  7087 &  K0 III &  4850 &  2.55  &  $-$0.15   &  &   4915   &  2.74  & $-$0.12 \\
  7318 &  K0 III &  4815 &  2.55  &  $-$0.11   &  &   4832   &  2.50  & $-$0.08 \\
 10380 &  K3 III &  4300 &  2.20  &  $-$0.27   &  &   4410   &  1.99  &  0.00   \\
 10761 &  K0 III &  4952 &  2.43  &   0.00     &  &   5020   &  2.84  & $-$0.24 \\
 12929 &  K2 III &  4600 &  2.70  &  $-$0.13   &  &   4480   &  2.28  & $-$0.04 \\
 13468 &  G9 III &  4893 &  2.54  &  $-$0.12   &  &   4919   &  2.64  & $-$0.42 \\
 15176 &  K1 III &  4650 &  2.85  &  $-$0.07   &  &   4531   &  2.45  & $-$0.08 \\
 18449 &  K2 III &  4500 &  2.65  &  $-$0.07   &  &   4314   &  2.30  &  0.00   \\
 19656 &  K1 III &  4600 &  2.30  &  $-$0.18   &  &   4726   &  2.46  &  0.00   \\
 21755 &  G8 III &  5012 &  2.45  &  $-$0.03   &  &   5064   &  2.95  & $-$0.38 \\
 26162 &  K2 III &  4800 &  2.90  &   0.06     &  &   4728   &  2.76  &  0.00   \\
 27382 &  K1 III &  4550 &  2.50  &  $-$0.32   &  &   4371   &  1.67  & $-$0.46 \\
 27697 &  G8 III &  4984 &  2.64  &   0.09     &  &   4955   &  3.06  &  0.00   \\
 28100 &  G8 III &  5011 &  2.54  &  $-$0.24   &  &   5039   &  3.17  & $-$0.33 \\
 28305 &  K0 III &  4883 &  2.57  &   0.05     &  &   4976   &  3.06  &  0.00   \\
 34559 &  G8 III &  4998 &  2.74  &   0.03     &  &   5084   &  3.20  & $-$0.12 \\
 38527 &  G8 III &  5046 &  2.77  &  $-$0.07   &  &   5122   &  3.29  & $-$0.29 \\
 42398 &  K0 III &  4650 &  2.40  &  $-$0.15   &  &   4646   &  2.51  & $-$0.17 \\
 48433 &  K1 III &  4550 &  2.20  &  $-$0.20   &  &   4528   &  2.15  &  0.00   \\
 52556 &  K1 III &  4700 &  2.65  &  $-$0.08   &  &   4627   &  2.55  &  0.00   \\
 54079 &  K0 III &  4450 &  2.10  &  $-$0.42   &  &   4415   &  1.85  & $-$0.38 \\
 54719 &  K2 III &  4500 &  2.55  &   0.14     &  &   4512   &  2.35  &  0.29   \\
 55751 &  K0 III &  4550 &  2.10  &  $-$0.11   &  &   4619   &  2.28  &  0.00   \\
 59686 &  K2 III &  4650 &  2.75  &   0.15     &  &   4706   &  2.51  &  0.00   \\
 65695 &  K2 III &  4470 &  2.45  &  $-$0.15   &  &   4436   &  2.14  &  0.00   \\
 69994 &  K1 III &  4650 &  2.60  &  $-$0.07   &  &   4518   &  2.38  & $-$0.08 \\
 73471 &  K2 III &  4550 &  2.40  &   0.05     &  &   4603   &  2.66  &  0.29   \\
184406 &  K3 III &  4520 &  2.41  &   0.04     &  &   4356   &  2.36  &  0.06   \\
190327 &  K0 III &  4850 &  2.70  &  $-$0.15   &  &   4832   &  2.83  &  0.18   \\
192944 &  G8 III &  5000 &  2.70  &  $-$0.10   &  &   5040   &  2.84  & $-$0.08 \\
194317 &  K3 III &  4435 &  2.70  &   0.04     &  &   4338   &  2.73  &  0.12   \\
197139 &  K2 III &  4485 &  2.40  &  $-$0.08   &  &   4439   &  1.87  & $-$0.04 \\
197912 &  K0 III &  4940 &  3.17  &  $-$0.03   &  &   4844   &  2.77  & $-$0.06 \\
197989 &  K0 III &  4785 &  2.27  &  $-$0.11   &  &   4791   &  2.25  & $-$0.04 \\
199253 &  K0 III &  4625 &  2.35  &  $-$0.19   &  &   4520   &  2.49  &  0.00   \\
209761 &  K2 III &  4420 &  2.35  &  $-$0.08   &  &   4371   &  1.98  &  0.00   \\
210762 &  K0 III &  4185 &  1.65  &   0.00     &  &   4181   &  1.98  &  0.42   \\
211388 &  K3 III &  4260 &  2.15  &   0.01     &  &   4173   &  1.48  &  0.46   \\
213119 &  K5 III &  3910 &  1.59  &  $-$0.48   &  &   3955   &  1.34  & $-$0.37 \\
214995 &  K0 III &  4680 &  2.70  &  $-$0.04   &  &   4540   &  2.25  &  0.13   \\
219615 &  G7 III &  4830 &  2.57  &  $-$0.54   &  &   4958   &  2.26  & $-$0.50 \\
220954 &  K1 III &  4775 &  2.95  &   0.02     &  &   4779   &  2.51  &  0.00   \\
223252 &  G8 III &  5031 &  2.72  &  $-$0.08   &  &   5041   &  2.88  & $-$0.12 \\
224533 &  G9 III &  5030 &  2.72  &  $-$0.02   &  &   5084   &  2.87  & $-$0.04 \\  
\multicolumn{9}{c}{}\\
\hline      
\end{tabular}
\end{center} 
\end{table}

\subsection{Tests with multi-epoch observation}

As noted in Sect. 3, 60 of the 245 target stars were observed at a
second epoch, typically at least three months apart.  The comparison of
first and second epoch observations offers another test of the accuracy and
consistency of the radial velocity provided in the catalog associated to
this paper.

The comparison is presented in Figure~3, which indicates a negligible value
for the mean difference between the velocities at the two epochs and a
standard deviation of the differences amounting to 2.6 km~s$^{-1}$.  

It is interesting to note that none of the 60 stars in Figure~3 shows a
radial velocity difference between first and second epoch exceeding
3$\sigma$, indicating that all observed differences could be entirely
ascribed to only observational uncertainties, without necessarily invoking
binarity or pulsation.
 
\section{Atmospheric parameters}

We derived the atmospheric parameters ($T_{\rm eff}$, $\log g$, [M/H])
of program stars via $\chi^2$ fitting to the synthetic spectral library of
Munari et al.  (2005), the same library as used in the analysis of RAVE
spectra, which is based on the atmospheric models of Castelli \& Kurucz
(2003).  We used the library in its solar scaled ($[\alpha/Fe]=0$) and 2
km~s$^{-1}$ micro-turbulence version, computed at a resolving power 5500,
the same as our observed spectra.

\subsection{Selecting the wavelength range for $\chi^2$}

We carried out extensive preliminary tests to investigate what should be the
optimal extension and positioning of the wavelength interval within the
recorded wavelength range over which to carry out the $\chi^2$ fitting.

The necessity of such a preliminary assessment was dictated by some
considerations.  On one side, the wider the wavelength range subject to
$\chi^2$, the larger the amount of information brought in.  On the other
hand, the wider the wavelength range, the less trustworthy is the continuum
normalization.  In addition, while the $\chi^2$ fitting is mainly driven
by the countless amount of weak, optically thin lines forming over a large
extension of the stellar atmosphere, the classical line-by-line analysis
(usually based on ionization and excitation balance of Fe lines) works on a
selected and limited sample of lines that could probe a less extended depth
of the stellar atmosphere.  Furthermore, while $\chi^2$ fitting is a far
more efficient method with large volumes of data and/or with medium (or low)
resolution spectra that are packed with blended lines, it is also far more
sensitive than the line-by-line method to the limited completeness of
the line lists used to compute the synthetic spectra (see for example  the huge
number of {\it new} FeII lines introduced by Castelli and Kurucz 2010).

For these and other subtler reasons we extensively explored what could be
gained by considering only a sub-interval of the whole wavelength range
covered by our spectra.  

To this aim, we used the spectra of the 87 RC stars that we selected from the
extensive surveys of RC stars by Hekker \& Melendez (2007), Takeda et al. 
(2008) and Soubiran \& Girard  (2005), and which we re-observed with our
instrument with the same S/N of the spectra collected for the target stars
(see next section for details).  The atmospheric analysis in these papers was
carried out on high-resolution spectra with the line-by-line method.  We
considered wavelength intervals progressively shorter (in steps of 100
\AA), and let them reposition freely within the recorded wavelength range
of our spectra.  For each spectrum and for each length and position of the
wavelength interval we compared the result of the $\chi^2$ fitting with that
of the line-by-line method.

The results of the comparison showed that the accuracy of the $\chi^2$
fitting does not improve by considering wavelength intervals wider than 300
\AA.  Next, we explored what would be the best positioning of a
$\Delta$$\lambda$=300 \AA\ interval that would minimize the difference
between the results of $\chi^2$ fitting and literature data based on the
line-by-line method.  We found that the interval 4761$-$5061~\AA\ delivered
the best results (in the sense of null off-set and lower dispersion) for
[M/H], 5614$-$5918~\AA\ for $\log g$, and 5012$-$5312~\AA\ for $T_{\rm eff}$.  
These intervals were consequently adopted for the analysis of target stars.

\begin{table}
\caption{Comparison between the atmospheric parameters obtained with
the $\chi^2$ method and those derived by Takeda et al.  (2008)}
\begin{center}
\begin{tabular}{@{~~~}l @{~~~}c @{~~~}c @{~~~}c @{~~~}r @{~~~}c @{~~~}c @{~~~}c @{~~~}r}
\hline
\multicolumn{9}{c}{}\\
 HD    & Spc.T. &\multicolumn{3}{c}{Takeda}     & &  \multicolumn{3}{c}{$\chi^2$}  \\ \cline{3-5} \cline{7-9} 
       &        & $T_{\rm eff}$ & log\textit{g} & [M/H] &    & $T_{\rm eff}$ & log\textit{g} & [M/H] \\ 
       &        &     (K)       &    (dex)      & (dex) &    &     (K)       &   (dex)       & (dex) \\
\multicolumn{9}{c}{}\\          
   448 & G9 III &  4780   &  2.51  &    0.03 & &     4731    &     2.63   &     0.04   \\
   587 & K1 III &  4893   &  3.08  & $-$0.09 & &     4791    &     3.25   &  $-$0.34   \\
  4627 & G8 III &  4599   &  2.05  & $-$0.20 & &     4664    &     2.42   &  $-$0.04   \\
  6186 & K0 III &  4829   &  2.30  & $-$0.31 & &     4845    &     2.55   &  $-$0.50   \\
  7087 & K0 III &  4850   &  2.55  & $-$0.04 & &     4915    &     2.74   &  $-$0.12   \\
 10348 & K0 III &  4931   &  2.55  &    0.01 & &     4895    &     3.06   &     0.00   \\
 10761 & K0 III &  4952   &  2.43  & $-$0.05 & &     5020    &     2.84   &  $-$0.24   \\
 10975 & K0 III &  4866   &  2.47  & $-$0.17 & &     4874    &     2.68   &  $-$0.25   \\
 11037 & G9 III &  4862   &  2.45  & $-$0.14 & &     4957    &     2.51   &  $-$0.37   \\
 13468 & G9 III &  4893   &  2.54  & $-$0.16 & &     4919    &     2.64   &  $-$0.42   \\
 19525 & G9 III &  4801   &  2.59  & $-$0.11 & &     4813    &     2.69   &  $-$0.08   \\
 21755 & G8 III &  5012   &  2.45  & $-$0.13 & &     5064    &     2.95   &  $-$0.38   \\
 23526 & G9 III &  4837   &  2.50  & $-$0.15 & &     4852    &     2.63   &  $-$0.42   \\
 27371 & G8 III &  4923   &  2.57  &    0.10 & &     4985    &     3.21   &     0.00   \\
 27697 & G8 III &  4984   &  2.64  &    0.12 & &     4955    &     3.06   &     0.00   \\
 28100 & G8 III &  5011   &  2.54  & $-$0.08 & &     5039    &     3.17   &  $-$0.33   \\
 28305 & K0 III &  4883   &  2.57  &    0.13 & &     4976    &     3.05   &     0.00   \\
 34559 & G8 III &  4998   &  2.74  &    0.00 & &     5084    &     3.20   &  $-$0.12   \\
 35410 & K0 III &  4809   &  2.58  & $-$0.33 & &     4794    &     2.94   &  $-$0.50   \\
 38527 & G8 III &  5046   &  2.77  & $-$0.11 & &     5122    &     3.29   &  $-$0.29   \\
 39007 & G8 III &  4994   &  2.69  &    0.08 & &     5148    &     3.30   &     0.00   \\
 39019 & G9 III &  4964   &  2.91  &    0.19 & &     5017    &     3.25   &     0.00   \\
 45415 & G9 III &  4753   &  2.39  & $-$0.12 & &     4852    &     2.73   &  $-$0.25   \\
 51814 & G8 III &  4846   &  2.23  & $-$0.02 & &     4753    &     2.79   &  $-$0.04   \\
192787 & K0 III &  5025   &  2.86  & $-$0.07 & &     4958    &     2.76   &  $-$0.12   \\
192944 & G8 III &  4981   &  2.48  & $-$0.06 & &     5040    &     2.84   &  $-$0.08   \\
209396 & K0 III &  4999   &  2.81  &    0.04 & &     5041    &     3.17   &  $-$0.04   \\
210434 & K0 III &  4949   &  2.93  &    0.12 & &     4919    &     3.13   &     0.08   \\
210702 & K1 III &  4967   &  3.19  &    0.01 & &     4974    &     3.21   &  $-$0.12   \\
217264 & K1 III &  4946   &  2.80  &    0.12 & &     4978    &     3.27   &     0.00   \\
219615 & G7 III &  4802   &  2.25  & $-$0.62 & &     4958    &     2.26   &  $-$0.50   \\
223252 & G8 III &  5031   &  2.72  & $-$0.03 & &     5041    &     2.88   &  $-$0.12   \\
224533 & G9 III &  5030   &  2.72  & $-$0.01 & &     5084    &     2.87   &  $-$0.04   \\
\multicolumn{9}{c}{}\\
\hline                                                                                       
\end{tabular}
\end{center}                                                                                 
\end{table}                                                                                  

\begin{table}
\caption{Comparison between the atmospheric parameters obtained with
the $\chi^2$ method and those derived by Soubiran \& Girard (2005)}
\begin{center}
\begin{tabular}{@{~~~}l @{~~~}l @{~~~}c @{~~~}c @{~~~}r @{~~~}c @{~~~}c @{~~~}c @{~~~}r}
\hline
\multicolumn{9}{c}{}\\
 HD    & Spc.T. &  \multicolumn{3}{c}{Soubiran \& Girardi} & &  \multicolumn{3}{c}{$\chi^2$}  \\ \cline{3-5} \cline{7-9} 
       &        & $T_{\rm eff}$ & log\textit{g} & [M/H] &    & $T_{\rm eff}$ & log\textit{g} & [M/H] \\ 
       &        &     (K)       &    (dex)      & (dex) &    &     (K)       &   (dex)       & (dex) \\
\multicolumn{9}{c}{}\\                                                                                           
188512 & G8 IV  &  5106  &  3.54  &   $-$0.15 & &     5073    &     3.42   &    $-$0.46   \\
190404 & K1 V   &  4963  &  3.90  &   $-$0.82 & &     4539    &     3.44   &    $-$0.66   \\
191026 & K0 IV  &  5050  &  3.49  &   $-$0.10 & &     4979    &     3.40   &    $-$0.25   \\                                                                                             
197964 & K1 IV  &  4813  &  3.03  &    0.17   & &     4773    &     3.13   &     0.46   \\ 
212943 & K0 III &  4586  &  2.81  &   $-$0.34 & &     4521    &     2.50   &    $-$0.29   \\
219615 & G7 III &  4830  &  2.57  &   $-$0.42 & &     4958    &     2.26   &    $-$0.50   \\                                                                                             
\multicolumn{9}{c}{}\\
\hline                                                                                       
\end{tabular}                                                                                
\end{center}                                                                                 
\end{table}  

\begin{figure}[htp]
\center
\includegraphics[totalheight=0.36\textheight, angle=270]{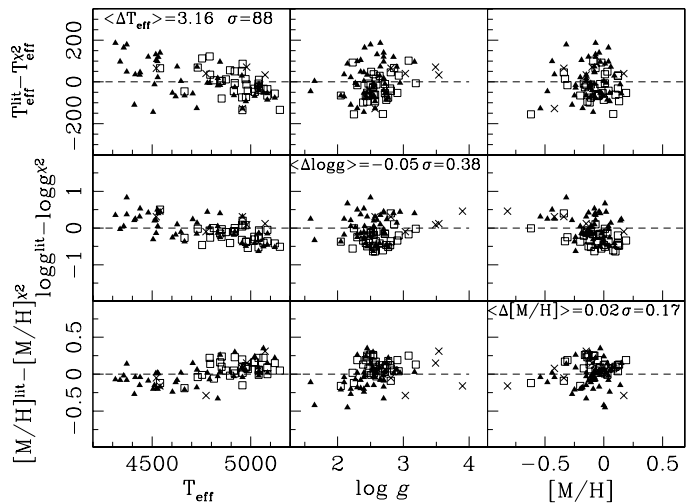}
\caption{Differences with literature for the atmospheric parameters derived 
from $\chi^2$ fitting for 87 red clump stars.  Empty squares are stars from 
Takeda et al. (2008), filled triangles Hekker \& Melendez (1997) and the 
crosses from Soubiran \& Girard (2005).}
\end{figure}

\begin{figure}[htp]
\center
\includegraphics[totalheight=0.36\textheight,angle=270]{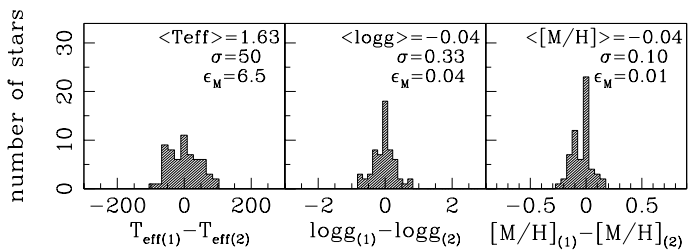}
\caption{The differences in effective temperature, gravity and metallicity
of repeated observations of 60 target stars.}
\end{figure}

\begin{figure}[htp]
\center
\includegraphics[totalheight=0.36\textheight, angle=270]{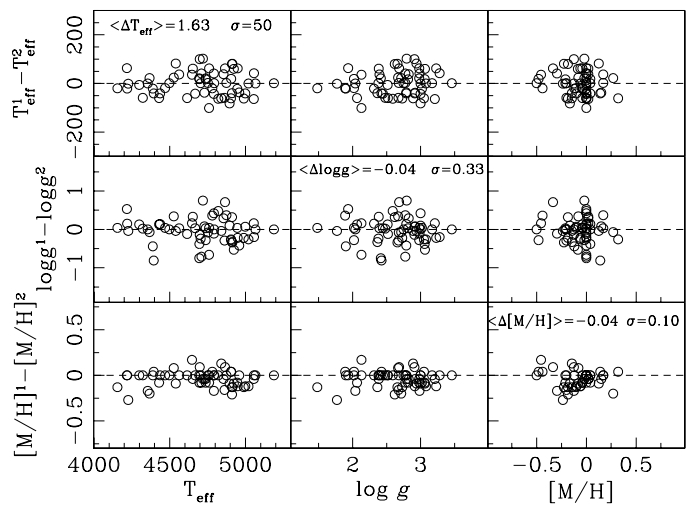}
\caption{Distribution of the differences in atmospheric parameters for
repeated observations of 60 RC stars .}
\end{figure}

\begin{table}
\caption{Content and description of the catalog}
\begin{center}
\scalebox{0.8}{
\begin{tabular}{@{~}r @{~}l @{~}l @{~}l @{~}l}
\hline
\multicolumn{5}{c}{}\\
Character & Code & Units & Symbol & Description \\
\hline
\multicolumn{5}{c}{}\\
1-8      &   I8    &  ...     &  HD      &  HD number   \\
9-15     &   I5    &  ...     &  HIP     &  HIP number  \\
16-21    &   I6    &  ...     &  TYC1    &  TYCHO-2 1st identi???er \\
25-30    &   I5    &  ...     &  TYC2    &  TYCHO-2 2nd identi???er \\
33       &   I1    &  ...     &  TYC3    &  TYCHO-2 3rd identi???er \\
37-45    &   A11   &  ...     &  spTyp   &  Spectral type from Michigan catalog \\
49-61    &   F12.8 &  ...     &  RA      &  Right ascension (J2000) \\
65-77    &   F12.8 &  ...     &  DE      &  Declination (J2000)  \\
82-90    &   F8.4  &  deg     &  GLat    &  Galactic latitude    \\
94-102   &   F8.4  &  deg     &  DLon    &  Galactic longitude   \\
106-112  &   F6.4  &  mag     &  Hp      &  Hipparcos Hp magnitude \\
114-120   &   F6.4  &  mag     &  eHp     &  error on Hp            \\
122-126  &   F4.2  &  mag     &  V-I     &  (V???I)C from Hipparcos catalog \\
130-134  &   F4.2  &  mag     &  eV-I    &  error on (V???I)C               \\
137-141  &   F7.2  &  mas     &  parHip  &  Hipparcos (ESA 1997) parallax \\
143-147  &   F6.2  &  mas     &  eparHip &  error on Hipparcos (ESA 1997) parallax \\
154-158  &   F7.2  &  mas     &  parVL   &  Hipparcos (van Leeuwen 2007) parallax   \\
163-167  &   F6.2  &  mas     &  eparVL  &  error Hipparcos (van Leeuwen 2007) parallax \\
169-175  &   F6.3  &  mag     &  BT      &  Tycho BT magnitude           \\
178-183  &   F5.3  &  mag     &  eBT     &  error on Tycho BT magnitude  \\
186-191  &   F6.3  &  mag     &  VT      &  Tycho VT magnitude           \\
194-199  &   F5.3  &  mag     &  eVT     &  error on Tycho VT magnitude  \\
201-207  &   F6.1  &  mas/yr  &  pmRA    &  Tycho-2 RA proper motion     \\
211-213  &   F4.1  &  mas/yr  &  epmRA   &  error on Tycho-2 RA proper motion \\
215-221  &   F6.1  &  mas/yr  &  pmDEC   &  Tycho-2 DEC proper motion         \\
223-226  &   F4.1  &  mas/yr  &  epmDEC  &  error on Tycho-2 DEC proper motion \\
230-235  &   F6.3  &  mag     &  J2MASS  &  2MASS J magnitude                  \\
242-247  &   F5.3  &  mag     &  eJ2MASS &  error on 2MASS J magnitude         \\
250-255  &   F6.3  &  mag     &  K2MASS  &  2MASS H magnitude                  \\
258-263  &   F5.3  &  mag     &  eK2MASS &  error on 2MASS H magnitude         \\
266-271  &   F6.3  &  mag     &  K2MASS  &  2MASS K magnitude                  \\
274-279  &   F5.3  &  mag     &  eK2MASS &  error on 2MASS K magnitude         \\
281-284  &   A3    &  ...     &  2MASSQF &  2MASS quality index                \\
286-291  &   F6.3  &  mag     &  IDENIS  &  DENIS I magnitude                  \\
298-302  &   F4.2  &  mag     &  eIDENIS &  error on IDENIS                    \\
305-308  &   I3    &  ...     &  DENISQF &  DENIS quality index                \\
311-327  &   F16.8 &  ...     &  HJD     &  Heliocentric Julian date of Observation \\
330-334  &   F4.2  &  ...     &  FWHM    & FWHM of the night sky lines         \\
336-341  &   F5.1  &  ...     &   S/N    & S/N ratio of the spectra            \\
343-349  &   F6.1  & (km s$^{-1}$) &  RV      &  Heliocentric radial velocity            \\
354-357  &   F3.1  & (km s$^{-1}$) &  eRV     &  Error on heliocentric radial velocity   \\
362-366  &   I5    &  K       &  Teff    &  Effective temperature                            \\
371-374  &   I2    &  K       &  eTeff   &  Error on effective temperature                   \\
379-383  &   F4.2  &  dex     &  logg    &  Surface gravity                                  \\
388-392  &   F5.2  &  dex     &  elogg   &  Error on surface gravity                         \\
398-403  &   F5.2  &  dex     &  [M/H]   &  Metallicity                                      \\
408-412  &   F4.2  &  dex     &  e[M/H]  &  Error on metallicity                             \\
419-424  &   F6.1  &  pc      &  d       &  Spectrophotometric distance                     \\
429-433  &   F5.1  &  pc      &  ed      &  Error on spectro-photometric distance            \\
437-445  &   F8.6  &  mag     &  Av      &  Extinction in $V$-band                           \\
448-457  &   F9.4  &  pc      &  X       &  Galactic X Coordinate                            \\
459-468  &   F9.4  &  pc      &  Y       &  Galactic Y coordinate                            \\
470-479  &   F9.4  &  pc      &  Z       &  Galactic Z coordinate                            \\
482-491  &   F8.4  & (km s$^{-1}$) &  U       &  U velocity                                       \\
492-502  &   F8.4  & (km s$^{-1}$) &  V       &  V velocity                                       \\
505-514  &   F8.4  & (km s$^{-1}$) &  W       &  W velocity                                       \\
\multicolumn{5}{c}{}\\
\hline                                                                                       \\
\end{tabular}}
\end{center}
\end{table}

\subsection{Comparison with literature data on red clump stars}
 
To test the accuracy of the results of our $\chi^2$ fitting, a comparison
was carried out against the results of existing surveys of RC stars.

Hekker \& Melendez (2007), Takeda et al. (2008) and
Soubiran \& Girard (2005) provided detailed chemical abundance
analyses of RC stars.  Based solely on observability at the telescope, we
selected 47 RC stars from Hekker \& Melendez and 34 RC stars from Takeda et
al., and observed them interspersed with the targets stars.  The exposure
time was adjusted so that the S/N of the spectra for these template stars
would match the median value of the S/N obtained from the target stars (to
avoid introducing biases, expecially during the continuum normalization
process). This sample of template stars was augmented by an additional six stars 
in the same spectral type range of RC stars, but of various luminosity classes, 
selected from Soubiran \& Girard.

The result of the comparison is given in Tables~2, 3, and 4, and presented in
graphical form in Figure 4. For the whole sample of 87 template stars, it is
\begin{eqnarray}
 <(T_{\rm our} - T_{\rm lit})> &=& +3 ~{\rm K} ~~~(\sigma=88 ~{\rm K})\\
 <(\log g_{\rm our} - \log g_{\rm lit})> &=& -0.05 ~~~(\sigma=0.38)\\
 <{\rm [M/H]}_{\rm our} - {\rm [M/H]}_{\rm lit}> &=& +0.02 ~~~(\sigma=0.17),
\end{eqnarray}
thus negligible zero-point offsets, and limited dispersions. The latter
could be even better if it were not inflated by the systematic differences between
the individual sources in literature.  Indeed considering the differences
with individual sources, for the 47 RC stars in common with Hekker \&
Melendez (2007) we have $<$($T_{\rm our}$$-$$T_{\rm HM}$)$>$= $-$11
K ($\sigma$=83 K), $<$($\log g_{\rm our}$$-$$\log g_{\rm HM}$)$>$ = $-$0.06
dex ($\sigma$=0.38 dex), $<$($[$M/H$]_{\rm our}$$-$$[$M/H$]_{\rm HM}$)$>$ =
0.04 ($\sigma$= 0.18), while the comparison with the 34 RC stars from Takeda
et al.  (2008) provides $<$($T_{\rm our}$$-$$T_{\rm Tak}$)$>$ =28 K
($\sigma$=61 K), $<$($\log g_{\rm our}$$-$$\log g_{\rm Tak}$)$>$=0.28
($\sigma$=0.23), $<$($[$M/H$]_{\rm our}$$-$$[$M/H$]_{\rm Tak}$)$>$=$-$0.10
($\sigma$= 0.11).  For the 147 stars in common among them, $T_{\rm eff}$ and
$\log g$ from Takeda et al.  (2008) are on average $\sim$50 K cooler and
$\sim$0.25 dex lower, respectively, than those of Hekker \& Melendez (2007). 
These differences are close to those we found in this paper, with Takeda et
al.  (2008) being 39 K cooler and 0.34 dex lighter in gravity than Hekker \&
Melendez (2007).

\subsection{Repeated observations}

As discussed in Sect.~3, we re-observed 60 target RC stars at a second
epoch, separated in time from the first observations by three months on average. 
Figure 5 presents the distribution of the differences in $T_{\rm eff}$,
$\log g$ and [M/H] between these two re-observations.  The distributions in
all three parameters are sharp and well behaving.  They are
characterized by $\sigma(T_{\rm eff})$=50 K, $\sigma(\log g)$=0.33 dex and
$\sigma$([M/H])=0.10 dex.  The behavior of the differences versus
$T_{\rm eff}$, log$_{g}$ are illustrated in Figure~6.  As for the observations
of the 87 template stars, the differences do not show a trend with the
atmospheric parameters.

\begin{figure}[htp]
\center
\includegraphics[totalheight=0.36\textheight, angle=270]{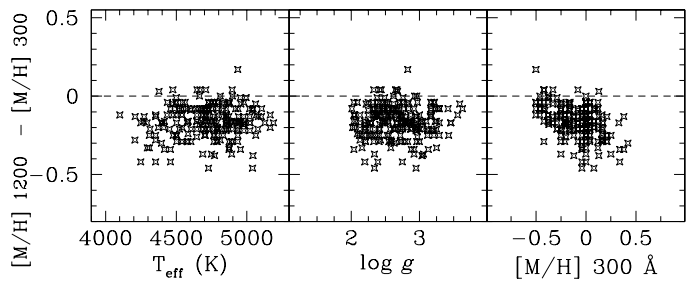}
\caption{Testing the difference in metallicity induced by widening the
wavelength range for $\chi^2$ fit from the adopted 300~\AA\ to
the 1200~\AA\ covered by untrimmed spectra.} 
\end{figure}

\subsection{The absence of an off-set in metallicity}

In Paper~I we found and corrected for a systematic off-set in metallicity
by $-$0.21 dex. The data in the present paper are not affected by a systematic
off-set, and therefore no correction was added to the $\chi^2$ results
when compiling the output catalog.

Both Paper I and this study used the same $\chi^2$ algorithm and the same
reference library of synthetic spectra, and thus some investigation was in
order to understand why the two papers show different results in terms of the
presence/absence of an off-set in metallicity.  To this aim, we carried out an
additional $\chi^2$ fit to the spectra discussed in this paper, this time
taking the whole 1200 \AA\ observed wavelength range, instead of the $\Delta
\lambda$=300~\AA\ adopted for the compilation of the annexed catalog.  The
untrimmed 1200 \AA\ range of our spectra is the same range of wavelengths
covered by the Echelle orders used in Paper I.  As remarked earlier, the
spectra extending over the whole 1200 \AA\ range show locally deep and
broad inflections owing to the presence of molecules, in particular MgH and
CN.  Guessing the true level of the continuum during the normalization
process is far more uncertain in these conditions than for the shorter, 300
\AA\ wide intervals used in this paper to derive the atmospheric parameters.

Figure~7 presents the difference in metallicity ($\Delta$[M/H]) of $\chi^2$
fits carried out over the whole 1200 \AA\ range and over the 300 \AA\
interval selected to measure the stellar metallicity.  As expected, there is
no systematics depending on $T_{\rm eff}$ and $\log g$ of the star, while a
clear one is present on metallicity, in the sense that the higher the
metallicity, the larger is the difference.  The mean value of the difference
is $-$0.16 dex, similar in sign and amount to the offset found for the
Echelle spectra of Paper I.  Extrapolating the linear trend visible in
Fig.~7, $\Delta$[M/H] is expected to null out at about [M/H]=$-$0.8.  The
interpretation seems straightforward.  The continuum normalization process
carried out unconstrained over the whole 1200 \AA\ range tends to slightly
fill in the molecular bands, causing the $\chi^2$ to react by returning a
reduced metallicity.  A lower metallicity weakens the intensity of the
molecular bands, and consequently the normalization process has less chance
to fill in these bands.  At [M/H]=$-$0.8 the molecular bands are
too weak to drive the normalization of the stellar continua off-road prior
to feeding the spectra to the $\chi^2$ fitting routine.

\section{The catalog}

The results of our observations are presented in the output catalog
(available electronically via CDS).  We tried to make it as similar as
possible to that accompanying Paper I.  The major difference is that we do
not provide a projected rotation velocity, given the lower resolution of our
spectra.  The catalog is divided into two parts, with observations for all
245 target stars going into the first part, and the second reporting the
results of the re-observations for 60 targets.  The content of the catalog
is given in Table~5.

In addition to photometric and astrometric data from the literature, the
catalog provides the distance, the U,V,W space velocities and the X,Y,Z
galacto-centric coordinates of the target stars.  These values were computed
exactly as in Paper I, where full details are provided.  In essence, we
derived uniform spectrophotometric distances for all our program stars,
adopting the intrinsic absolute magnitudes in the Johnson $V$ band
calibrated by Keenan \& Barnbaum (2000).  These absolute magnitudes are
given separately for the various spectral types (G8III to K2III) covered by
our program stars, and are calibrated on RC stars with the most precise
Hipparcos parallaxes.  To derive the distances, we transformed the Tycho-2
$V_{\rm T}$ into the corresponding Johnson $V$ band following Bessell (2000)
relations, and adopted the (always very low) reddening derived from the
all-sky 3D mapping by Arenou et al.  (1992) and Drimmel et al.  (2003).  The
$U$, $V$, $W$ velocities were derived from position ($\alpha$,
$\delta$), proper motion ($\mu_{\alpha}$, $\mu_{\delta}$), radial velocity,
and distance following the formalism of Johnson and Soderblom (1987).  We
adopted a left-handed system with U becoming positive outward from the
Galactic center.  We adopted for the Sun velocity components with respect to
the LSR the values found by Dehnen \& Binney (1998, in km~sec$^{-1}$):
$U_\odot$=$-$10.0, $V_\odot$=5.23, $W_\odot$=7.17.

\begin{acknowledgements}            

We would like to thank A. Siviero, M. Valentini, K. Freeman, and R.  Barbon
for their help; and P.  Rafanelli, S.  Ciroi, S.  Di Mille and M.  Fiaschi
for supporting and encouraging observations with the Asiago telescopes.  TS
has been supported by ELSA (European Leadership in Space Astrometry) Marie
Curie grant under FP6 contract MRTN-CT-2006-033481.

\end{acknowledgements}

\end{document}